\documentstyle[12pt,epsf]{article}
\begin{document}
\tolerance=5000
\def\be{\begin{equation}}
\def\ee{\end{equation}}
\def\bea{\begin{eqnarray}}
\def\eea{\end{eqnarray}}
\def\nn{\nonumber \\}
\def\cF{{\cal F}}
\def\det{{\rm det\,}}
\def\Tr{{\rm Tr\,}}
\def\e{{\rm e}}
\def\etal{{\it et al.}}
\def\erp2{{\rm e}^{2\rho}}
\def\erm2{{\rm e}^{-2\rho}}
\def\er4{{\rm e}^{4\rho}}
\def\etal{{\it et al.}}
\def\D{{D \hskip -3mm /\,}}

\begin{center}
{\bf\large  Unified approach to study quantum properties of 
primordial black holes, wormholes and of quantum cosmology}

\ 

{\sc S. Nojiri}\footnote{
e-mail: nojiri@cc.nda.ac.jp}, 
{\sc O. Obregon$^{\clubsuit}$}\footnote{e-mail: 
octavio@ifug3.ugto.mx} and 
{\sc S.D. Odintsov$^{\spadesuit}$}\footnote{
e-mail: odintsov@mail.tomsknet.ru, odintsov@tspu.edu.ru} 

\ 

{\sl Department of Mathematics and Physics, National Defence Academy, 
Hashirimizu Yokosuka 239, JAPAN}

{\sl $\spadesuit$ 
Tomsk State Pedagogical University, 634041 Tomsk, RUSSIA}

{\sl $\clubsuit$
Instituto de Fisica de la Universidad de Guanajuato \\
P.O. Box E-143, 37150 Leon Gto., Mexico}

\ 

{\bf Abstract}

\end{center}

{\small 
We review the anomaly induced effective action for 
dilaton coupled spinors and scalars in large $N$ and $s$-wave 
approximation. It may be applied to study the following fundamental 
problems: construction of quantum corrected black holes (BHs),
inducing of primordial wormholes in the early Universe 
(this effect is confirmed)  
and the solution of initial singularity problem.
The recently discovered anti-evaporation of 
multiple horizon BHs is discussed. The existance 
of such primordial BHs may be interpreted as SUSY manifestation. 
Quantum corrections to BHs thermodynamics maybe 
also discussed within such scheme.
}

\ 

It is expected that quantum field theory (or more exactly 
quantum gravity) should be extremely important in the study 
of early Universe and black hole physics. There are different 
ways to incorporate some ``traces" of quantum gravity to
 cosmology. One very promising way is related with 
the effective action approach \cite{BOS}. According to
effective action formalism one can find quantum corrections 
in the form of some (non-local) functional which should be added 
to the action of classical gravity. Then, the final theory may be 
considered as some ``new" effective theory of gravity which should 
be discussed basically only as classical theory.
Unfortunately, there is strong barrier which stops the
realization of this beautiful scenario. That is the fact that 
in general it is impossible to calculate effective action in 
closed form. 
One can find it only as some expansion (usually on powers 
of curvature) or yet to calculate it exactly on some background 
with high symmetry (say de Sitter space).

Recently, there has been some progress in the calculation
of effective action for arbitrary space in closed form 
using $s$-wave\footnote{This approximation works for 
smooth enough backgrounds when one can neglect higher 
derivative terms in EA \cite{SD}.} and large $N$ approximation. 
The corresponding effective action 
may be applied to the study of circle of phenomena in black hole 
physics or early Universe in the situations where typical 
curvatures are strong enough. In this brief review, we discuss the 
application (in unified form) of such effective action
to the following problems:
\begin{enumerate}
\item Anti-evaporation of multiple horizon BHs due to
quantum effects.
\item Induction of primordial wormholes in the early Universe.
\item Solution of initial singularity problem.
\item Quantum corrections to BH thermodynamics.
\end{enumerate}
As classical action we consider standard Einstein action
\be
\label{gi}
S=-{1 \over 16\pi G}\int d^4x \sqrt{-g_{(4)}}
\left\{R^{(4)} -2\Lambda\right\}\ . 
\ee
In order to find effective action (back-reaction) to
above classical theory, we consider $N$ real minimal scalars and $M$ 
Majorana spinors whose action is given by
\be
\label{gii}
S= \int d^4x\,\sqrt{-g_{(4)}}\,
\left(\frac{1}{2}\sum_{i=1}^Ng^{\alpha\beta}_{(4)}
\partial_\alpha \chi_i \partial_\beta \chi_i + \sum_{i=1}^M 
\bar\psi_i \gamma^\mu\nabla_\mu \psi_i \right)\ .
\ee
The anomaly induced effective action for the above theory
can be evaluated for $M=0$ case \cite{2,3}\footnote{ For the study
of conformal anomaly for dilaton coupled scalars, see refs.\cite{EE,2,3}.}
 and $M\neq 0$ case 
\cite{NNO} using large $N$ and $s$-wave approximation.
Large $N$ approximation is necessary in order to 
justify why we do not consider gravity quantum contributions 
(which may be nevertheless discussed afterwards).
$s$-wave reduction makes the problem solvable on analytical level 
and it looks quite natural as we discuss below only metrics
with spherical symmetry, i.e. of the sort 
$ds^2=g_{\mu\nu}dx^\mu dx^\nu + \e^{-2\phi} d\Omega$. 
Here $\mu,\nu=0,1$, $g_{\mu\nu}$ and $\phi$ depend only 
on $x^0=t$, $x^1=r$. By $s$-wave reduction we mean 
that in quantum calculations we neglect the dependence 
from two spherical coordinates.

>From 2d conformal anomaly for dilaton coupled scalar and dilaton 
coupled spinor, one can find the anomaly induced 
effective action. It can be written in the following form 
\cite{3,6,NNO}:
\bea
\label{giv}
W&=&-{1 \over 8\pi}\int d^2x \sqrt{-g}\,\left[
{N+M \over 12}R{1 \over \Delta}R - N \nabla^\lambda \phi
\nabla_\lambda \phi {1 \over \Delta}R \right. \nn
&& \left. +\left(N + {2M \over 3} \right)\phi R
+2N\ln\tilde\mu^2 \nabla^\lambda \phi \nabla_\lambda \phi \right]\ .
\eea
Solving the equations of motion which follow from the action $S+W$, 
Bousso and Hawking \cite{BH2} have found a special solution 
describing the quantum corrected Schwarzschild-de Sitter
(or in nearly degenerated case called also as Nariai) 
space \cite{Nariai}. On 
the classical level SdS BH looks as follows:
\be
\label{sds}
ds^2=-V(r)dt^2 + V(r)^{-1}dr^2 + r^2d\Omega^2\ ,
\ \ V(r)=1-{2 \mu \over r} - {\Lambda \over 3}r^2\ .
\ee
Here $\mu$ is the black hole mass. 
$V(r)$ has two positive roots $r_c$ and $r_b$ ($r_c>r_b$), which 
correspond to cosmological and black hole horizons. The Nariai space 
is given as a limit of $r_c\rightarrow r_b$. By defining new 
time and radial coordinates
$t={1 \over \epsilon\sqrt\Lambda}\psi$ and 
$r={1 \over \sqrt\Lambda}\left[1-\epsilon \cos\chi 
-{\epsilon^2 \over 6}\right]$ (here $\epsilon$ is defined by 
$9\mu^2\Lambda=1 - 3\epsilon^2$), we obtain the metric in the 
nearly degenerate limit:
\bea
\label{sds2}
ds^2&=&-{1 \over \Lambda}\left(1 + {2\epsilon \over 3}\cos\chi
\right)\sin^2\chi d\psi^2 + {1 \over \Lambda}\left(1 
- {2\epsilon \over 3}\cos\chi\right)d\chi^2 \nn
&& +{1 \over \Lambda}\left(1 - 2\epsilon \cos\chi
\right)d\Omega^2\ .
\eea
The Nariai limit corresponds to $\epsilon\rightarrow 0$. Since 
$\chi$ has a periodiciy of $2\pi$, the topology of the Nariai 
limit could be $S_1\times S_2$. 
In the quantum level, the parameters of this metric are changed 
due to the influence of effective action part\cite{NNO}.
However, the structure of the metric is the same.

At the next step\cite{NO}, the authors have investigated the 
backreaction from $N$ conformal scalars $\chi_i$, $N_1$ vectors 
$A_\mu$ and $N_{1/2}$ Dirac spinors $\psi_i$ to Nariai metric. 
The corresponding effective action may be calculated 
even beyond  $s$-wave approximation. 
It was demonstrated in \cite{BH2}, that a nearly degenerated Nariai 
black hole may not only evaporate but also anti-evaporate (i.e. 
its radius is getting to grow due to quantum effects). 
In \cite{NO}, it has been shown that the Nariai solution is stable 
and anti-evaporation can occur only if GUT matter content satisfies 
\be 
\label{gx} 
2N+7N_{1/2}>26 N_1 . 
\ee 
This relation holds for $SU(5)$ GUT.
In \cite{ENO}, the case of the usual SO(10) GUT was investigated. 
Although Eq. (\ref{gx}) cannot be satisfied in the case of 
the non-supersymmetric model, Eq. (\ref{gx}) can be satisfied in 
the case of the supersymmetric SO(10) GUTs due to the contribution 
from the Higgsino in the large dimensional representations. Hence, 
one extremely bright application of effective action is the discovery 
of the process which is opposite to famous Hawking radiation 
and which was called the anti-evaporation of multiple horizon 
black holes. As a result, we may expect that not all primordial 
SdS black holes evaporated quickly in early Universe. 
Some of them could existed for a longer time (until the 
anti-evaporation has been stopped by some other processes) and even 
could be searched in the present Universe. Moreover, their existence 
may be interpreted as "supersymmetry test" as anti-evaporation 
may occur presumbly for SUSY GUTs.

Another related research may be connected with the possibilities 
to induce primordial wormholes at the early Universe. 
The corresponding metric is 
\be
\label{e10}
ds^2=-\e^{2\rho}dt^2 + dl^2 + \e^{-2\phi}d\Omega^2 ,
\ee
where $\rho=\rho(l)$, $\phi=\phi(l)$ and $l$ is the proper distance. 
We call $f(l)=\exp (2\rho)$ as redshift function and 
$r(l)=\exp (-\phi)$ as shape function.
This metric has again spherical symmetry. 
Note that the above wormhole cannot occur as 
solution of classical gravity equations with 
usual non-quantum matter due to severe limitations.
 
Applying the same effective action (\ref{giv}), 
we made the numerical study for some contents of matter.
The results of this study \cite{NOOO} are given at figures, for
the shape function $r(l)$ in fig.1 and for 
the redshift function $f(l)$ in fig.2. As we see, one gets  
the wormhole solution with increasing throat radius and 
increasing red-shift function. That proves the possibility 
of quantum inducing of wormholes at the early Universe. Hence 
quantum effects give rise the primordial wormholes at the 
very early Universe. This fact has been also confirmed very 
recently for $N=4$ super Yang-Mills theory in 
curved spacetime (via application of 4d anomaly induced 
effective action) in ref.\cite{OO}.

The effective action (\ref{giv}) can also be used to investigate 
quantum cosmology. In this way one can try to
find the answer to the problem 
of initial singularity.

\epsffile[0 320 500 550]{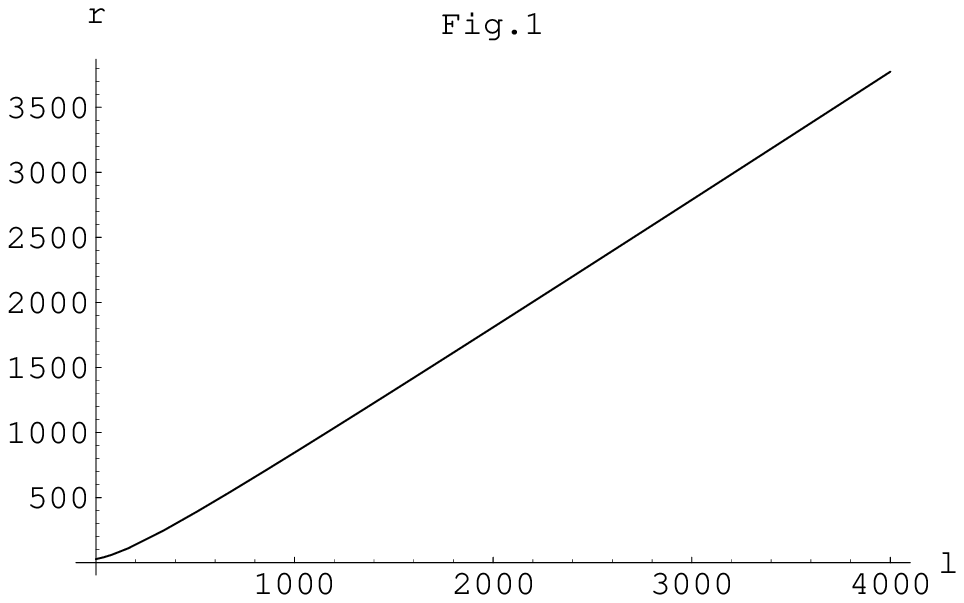}

\epsffile[0 320 500 550]{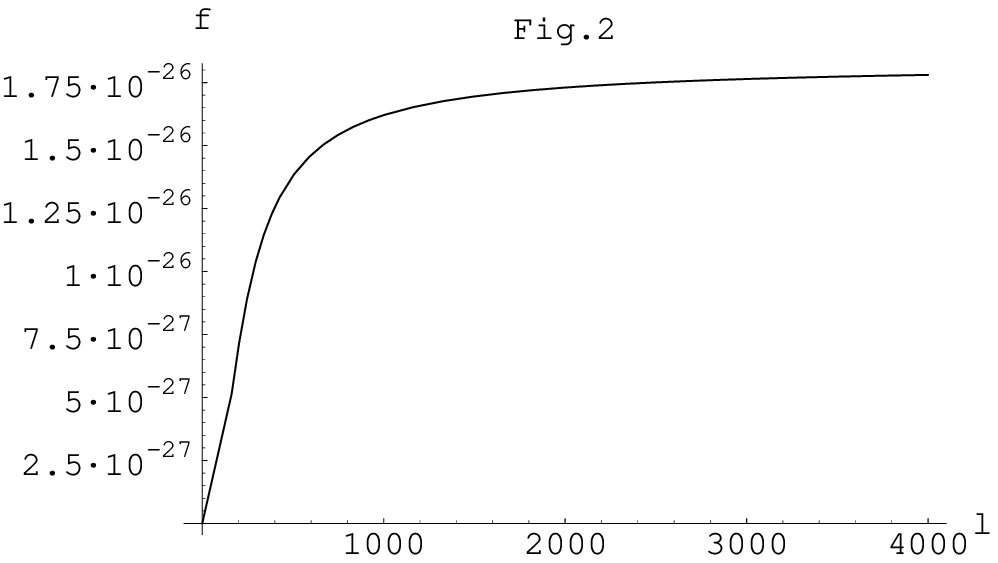}

When we assume that our spherically symmetric metric
does not depend on the 
radial coordinate $r$ but on the time coordinate $t$ only and $r$ is 
a periodic coordinate, there was found the
solution of effective equations of motion. The corresponding 
metric describes the Kantowski-Sachs Universe \cite{KS} 
whose topology is given by $S_1\times S_2$ \cite{11,NOOO2}. 
Especially in \cite{NOOO2}, an analytical solution without 
curvature singularity is found: 
\be
\label{Nvii}
ds^2= - d\tau^2 + {2 \over R_0}\cosh^2\left(\tau\sqrt{R_0 \over 2}
\right)dr^2 + \e^{-2\phi_0}d\Omega^2\ .
\ee
Here the dilaton field $\phi$ and the 2d scalar curvature $R$ 
become constants $\phi=\phi_0$ and $R=R_0$, which are given by
\bea
\label{Niii}
\e^{-2\phi_0}&\equiv& {(2N+M)G \over 6} + {1 \over 2\Lambda}
\pm {1 \over 2}\sqrt{{(2N+M)^2G^2 \over 9}+{1 \over \Lambda^2}
- {(8N + 6M)G \over 3\Lambda}} \nn
R_0&=&-{3\Lambda \over (N+M)G}\left({(2N+M)G \over 3} 
- {1 \over \Lambda} \right. \nn && \left.
\pm \sqrt{{(2N+M)^2G^2 \over 9}+{1 \over \Lambda^2}
- {(8N + 6M)G \over 3\Lambda}}\right)\ .
\eea
This KS Universe is expanding with the radius which is never zero. 
It may be completely induced by quantum effects (as it can be
considered in the regime where there is no classical 
corresponding solution). Moreover, it can presumbly  describe some 
sub-stage of inflationary Universe where there is effective expansion 
only along one of spatial coordinates. This explicit example 
opens a new way to re-consideration of initial singularity 
problem in frames of quantum gravity using effective action 
formalism.

As another application, we can consider the quantum corrections 
to the 4d Schwarzschild-(Anti-)de Sitter (S(A)dS) black holes (BH), 
whose metric is given by 
\bea
\label{SadSi}
&& ds^2 = - \e^{2\rho} dt^2 + \e^{-2\rho}dr^2
+ r^2 d\Omega^2 \nn
&& \e^{2\rho}=1 - {\mu \over x} - {\Lambda \over 3}x^2 
=-{\Lambda \over 3x}\prod_{i=1}^3 (x - x_i)\ .
\eea
Here $\mu$ is a constant of the integration corresponding to 
the black hole mass ($\mu =2GM_{\rm BH}$). 
The parameters $x_i$ ($i=1,2,3$) 
are solutions of the equation $\e^{2\rho_0}=0$. Among $x_i$'s, 
two are real and positive if 
$\Lambda>0$ and $\mu^2 < {4 \over 9\Lambda}$ 
and they correspond to black hole and cosmological horizons 
in the Schwarzschild-de Sitter black hole. On the other hand, only 
one is real if $\Lambda<0$. 

In Ref.\cite{NOthermo}, It has been found the temperature $T$
and entropy $S$ with account of quantum effects for multiply 
horizon SdS BH and SAdS BH as following
\bea
\label{SadSii}
T&\sim& \left| {\Lambda \over 12\pi}\left\{Y_I 
+\left(- 6 + {2Y_I \over x_I}\right)GN C_I
-GNY_IB_I \right\}\right| \nn
S&\sim&\pi x_I^2 - 4\pi GN C_I \\
B_I&\equiv & \Delta_0 + A\left[{7 \over 24x_I^2} 
- {1 \over 12}\sum_{i=1, i\neq I}^3\left\{
\left(1 - {Y_I^2 \over Y_i^2}\right)\left(  {1 \over x_i(x_I - x_i)}
- {1 \over x_i^2}\ln(x_I -x_i) \right)\right.\right. \nn
&& \left.\left. + {1 \over x_i^2 }\ln x_I- {2 \over x_ix_I} 
- {1 \over 6}\left(\sum_{j\neq i, j=1}^3 
{1 \over x_i(x_i - x_j)}\right)
\ln (x_I-x_i) 
\right\} - {1 \over 12x_I^2}\ln x_I\right] \nn
&& + {a' +B-1\over 2x_I^2} - {3 \over 4x_I^2} \nn
&& + \sum_{i=1, i\neq I}^3 \left\{{1 \over 2}\left({1 \over x_I^2}
-{1 \over x_i^2}\right)\ln (x_I-x_i)
+ {1 \over 2x_i^2}\ln x_I - {1 \over 2x_i x_I}\right\} \nn
C_I&\equiv& {1 \over \prod_{i=1,i\neq I}^3(x - x_i)} \nn
&& \times \left[ \Delta_1 + A \left\{ {x_I \over 6} 
- {X_2 \over 24x_I} + {7X_3 \over 48x_I^2} 
- {1 \over 24}\sum_{i=1, i\neq I}^3\left(Y_i - 
{Y_I^2 \over Y_i}\right)\ln\left(1 - {x_i \over x_I}
\right) \right\} \right. \nn
&& +{a'-B+1 \over 2}\left( x_I - {X_2 \over x_I} 
+ {X_3 \over 2x_I^2}\right) 
+ B\left(x_I + {X_3 \over 4x_I^2}\right) \nn
&& - {x_I \over 2}\left(\ln x_I - 1\right) 
+ {X_3 \over 4}\left(\ln x_I\right)^2 
+ {X_2 \over 2x_I}\left(\ln x_I + 1 \right) \nn
&& - {X_3 \over 4x_I^2}\left(\ln x_I + {1 \over 2}\right) 
+ {1 \over 2}\sum_{i=1,i\neq I}^3 
(x_I -x_i)\left(\ln (x_I -x_i) -1 \right) \nn
&& + {X_2 \over 2}\sum_{i=1,i\neq I}^3
\left( - {1 \over x_I}\ln (x_I - x_i)
+ {1 \over x_i}\ln\left(1 - {x_i \over x_I}\right)\right) 
- {X_2 \over 2x_I}\ln x_I \nn
&& - {X_3 \over 2}\left(-{1 \over 2x_I^2}
\ln \left( x_I - x_i \right)
+ {1 \over 2x_i^2}\ln\left(1 - {x_i \over x_I}\right)
+ {1 \over 2x_ix_I} \right) \nn
&& \left. - {X_3 \over 2}\left( -{1 \over 2x_I^2}\ln x_I 
+ {1 \over 2x_I^2}\right) \right]\ .\nonumber
\eea
Here
\bea
\label{SadSiii}
&& A\equiv {N+M \over N}\ , \quad 
B\equiv {N + {2M \over 3} \over N} \nn
&& Y_I\equiv {1 \over x_I}\prod_{i=1,i\neq I}^3 (x_i - x_I) \ ,
\eea
(\ref{SadSii}) also gives the corresponding 
expressions for their limits: Schwarzschild and de Sitter spaces. 
The latter can be regarded as the quantum correction to 
entropy of expanding Universe. 
Most of the above results are given 
for the same gravitational background with interpretation as 
4d quantum corrected BH or 2d quantum corrected dilatonic BH
\cite{6,MK}.

We presented Brief Review of some recent applications 
of anomaly induced effective action to various problems 
of quantum BHs physics and early Universe. There are many more topics 
where these methods maybe successfully applied. Let us mention
 only one which was discussed recently in refs.\cite{Bou,6} 
where it was shown that anti-evaporation of BHs in de Sitter 
Universe may lead to its proliferation. In other words, 
on the quantum level the fragmentation of de Sitter Universe to
few new daughter Universes may occur.

\

\noindent
{\bf Acknoweledgements} SDO would like to thank R. Bousso, S. Hawking 
and P. van Nieuwenhuizen for stimulating discussions. The work by OO
has been partially supported by  CONACyT Grants 3898P-E9608 and
28454-E.

\end{document}